# Virtual Reality for Large-Scale Laboratories Based on Colorized Point Clouds: design and pedagogical impact

**Lei Fan [1,*] and Yuxin Li [1,2]**

[1] Department of Civil Engineering, Xi'an Jiaotong-Liverpool University, Suzhou, 215123, China
[2] School of Engineering, University of Liverpool, Liverpool, L69 3BX, UK
* Correspondence: Corresponding author (Lei.Fan@xjtlu.edu.cn)

**Abstract**

Effective laboratory training is essential in engineering education, yet conventional on-site instruction is often constrained by time, accessibility, and safety considerations. To address these challenges, this study presents the design, implementation, and evaluation of a web-based virtual reality (WebVR) representation of a large-scale engineering laboratory constructed from massive colorized point cloud data. This study proposes a novel WebVR approach that integrates Unity and Potree for high-fidelity point-cloud visualization combined with advanced interactive capabilities in a browser-based virtual laboratory. It supports immersive first-person exploration, guided navigation, interactive hotspots conveying equipment and safety information, as well as emergency evacuation simulations. The usability, usefulness, and acceptance of the virtual laboratory were evaluated through an anonymous questionnaire administered to students and laboratory staff. User evaluation results indicated consistently positive feedback, with 100% of respondents rating the interface/navigation and visual/interactive content as good or excellent, 88.6% identifying scene realism as the biggest system strength (the most frequently selected), 74.3% reporting significantly higher engagement compared with traditional online laboratory training, and 82.9% indicating they would definitely recommend the system as a learning resource. In addition, a thematic analysis of qualitative feedback was performed to inform future enhancements of the WebVR environment. Overall, the findings demonstrate that the WebVR-based virtual laboratory can effectively complement conventional on-site laboratory instruction, offering a scalable, accessible, and low-risk platform that enhances learning experiences in engineering education.

## 1. Introduction

Effective engineering education extends beyond theoretical instruction and depends on sustained engagement with laboratory environments, where students can contextualize knowledge and develop practical competencies [1,2]. To facilitate laboratory activities, laboratory training is essential. Traditionally, laboratory training begins with in-person tours, safety briefings, and/or written documentation that introduce facility layouts, equipment usage, operational procedures, and emergency protocols.

Despite their widespread use, conventional training methods have persistent limitations. In-person tours are time-consuming, inflexible, and difficult to scale or adapt to the continuous arrival of new laboratory users. Written materials, meanwhile, often fail to convey three-dimensional spatial relationships, limiting learners' ability to form accurate mental maps of equipment, emergency exits, and evacuation routes. Traditional orientations are also typically one-time events, offering little opportunity for repetition, self-paced learning, or reinforcement. Consequently, knowledge retention declines quickly, leaving students and staff poorly prepared for emergencies. Prior studies report a lack of systematic training in laboratory risk awareness and safe operation [3], and identify inadequate understanding of laboratory environments and rule violations as contributing factors in university laboratory accidents [4,5].

In this context, immersive virtual reality (VR) offers a promising alternative for laboratory training. By immersing learners in interactive three-dimensional representations of real spaces, VR enables first-person exploration that closely mirrors physical presence [6], which is central to laboratory safety and facility training. Recent advances in VR hardware, software, and digital twin technologies make it feasible to construct high-fidelity virtual replicas of complex engineering laboratories, preserving accurate geometry, scale, and equipment layout. Crucially, such training can be conducted without physical access constraints or exposure to real-world risks, addressing long-standing limitations of traditional laboratory training [7].

Educational theory provides strong support for the use of immersive VR in laboratory training. Constructivist and experiential learning frameworks argue that knowledge is most effectively acquired through active, contextualized engagement rather than passive information delivery [8,9]. Immersive VR aligns with these principles by enabling learning through direct interaction with virtual artifacts, exploration of spatial environments, and immediate feedback on user actions. From a cognitive perspective, immersive environments can reduce extraneous cognitive load by directing attention to relevant features, enabling safe repetition, and fostering familiarity with complex or hazardous settings prior to physical entry [10].

Empirical studies further substantiate the effectiveness of VR-based laboratory training. Across science and engineering education, VR has been shown to produce learning outcomes comparable to or better than those achieved through traditional approaches. For instance, Tsirulnikov et al. [11] reported significantly higher post-test scores and increased motivation among students using VR laboratories, with 91% of participants considering VR a valuable educational supplement. Likewise, Levonis et al. [12] found that chemistry undergraduates unanimously perceived virtual laboratory tours as effective learning aids. Collectively, these findings indicate that VR can enhance laboratory comprehension, learner confidence, and engagement, outcomes that are especially important in safety-critical laboratory environments.

Despite these advances, a clear research gap persists. Most existing VR-based laboratory training systems target spatially well-organized environments, such as medical or chemical laboratories, where procedures are standardized and training tasks are clearly defined. In contrast, engineering laboratories typically involve heterogeneous equipment, complex spatial layouts, and distributed safety hazards, making comprehensive virtual representation substantially more challenging. To date, few studies have reported systems that integrate high-fidelity spatial reconstruction, safety-oriented interactivity, and orientation-focused pedagogy for large-scale engineering laboratories. As a result, both practical methodologies for developing such VR systems and empirical evidence of their effectiveness in supporting engineering laboratory safety awareness and facility familiarization remain limited.

To address this gap, this study presents the design, implementation, and evaluation of an immersive WebVR (web-browser-based virtual reality) laboratory aimed at enhancing safety awareness and facility familiarity in engineering education. The primary research question investigates how highly spatially accurate and detailed, yet efficiently rendered, web-based virtual environments of large-scale engineering laboratories can be developed from massive point cloud datasets. The secondary research question examines the users' perceived usability and usefulness of the established virtual laboratory environment.

Using a large civil engineering laboratory as a case study, a point-cloud-based virtual replica was developed through a hybrid Unity–Potree WebVR approach. The virtual environment incorporates interactive equipment labels, contextual information prompts, and clearly identified safety features, including first-aid stations, fire-fighting equipment, and evacuation routes. The system supports first-person navigation, measurement tools, and guided exploration, and is deployed as a lightweight, web-based application to maximize accessibility across devices. A user study involving undergraduate students, postgraduate students, and staff was conducted to evaluate usability, realism, engagement, and perceived educational value.

The main contributions of this work are threefold:

- Propose a novel hybrid WebVR approach that integrates Unity and Potree to enable immersive laboratory safety training and facility orientation within complex engineering environments.
- Establish a practical development workflow for constructing high-fidelity, efficiently rendered virtual environments of large-scale engineering laboratories using massive point cloud data.
- Provide empirical evaluation results on user perceptions of usability, usefulness, and acceptance of the developed WebVR-based laboratory, thereby demonstrating its potential as a scalable and effective immersive supplement to conventional on-site laboratory training.

## 2. Literature Review

### *2.1. Technology-Enhanced Learning*

With the limitations of traditional education increasingly evident, higher education is undergoing a digital transformation, where Technology-Enhanced Learning (TEL) has become a central theme [13]. TEL emphasizes using digital tools and technological means to optimize educational processes and improve learning experiences. However, the mere introduction of technology does not automatically improve educational quality; outcomes depend on factors such as learning environment, resources, and student engagement [14].

Student engagement can be measured using Reeve et al.'s four-dimensional framework: behavioral, emotional, cognitive, and active engagement [15-17]. Behavioral engagement, often limited in traditional teaching due to few opportunities for hands-on practice, can be enhanced through simulated exploration, supporting the experiential cycle of doing, reflecting, understanding, and applying knowledge [18]. Emotional engagement, critical for motivation, is facilitated by TEL tools satisfying learners' needs for autonomy, competence, and relatedness. Cognitive engagement, challenged by static materials, benefits from intuitive, structured, and dynamic visualizations, which reduce extraneous cognitive load and support integration of multiple sensory inputs such as visual, auditory, motor, and spatial information [19-22]. Active engagement, reflecting higher-order learning, aligns with constructivist theory: learners actively integrate new knowledge with experience through interaction and exploration, forming meaningful knowledge structures [23-25].

Beyond engagement, TEL is supported by broader educational informatization trends. Digital resources are easier to maintain and update than traditional materials, while dynamic management of learner behavior and teaching archives enables real-time analysis for instructional improvement [26,27]. Overall, TEL offers flexible, accessible, and sustainable learning environments, demonstrating potential to enhance educational outcomes from both macro and micro perspectives.

### *2.2. Virtual Reality Technology*

VR, originating in the 1960s, integrates computer graphics, sensors, and human-computer interfaces to provide immersive, three-dimensional experiences [28]. In education, VR can visualize abstract spatial structures, simulate complex scenarios, and create safe, risk-free environments for experiential learning. Spatial learning theory suggests that three-dimensional environments improve spatial perception, particularly in complex settings, while VR also allows repeated practice without physical or temporal constraints, enhancing long-term retention [21,29].

VR's immersive nature engages multiple senses, improving motivation, attention, and knowledge retention [30], while its interactive capabilities promote active, learner-centered exploration, aligning with "learning by doing". Importantly, the four characteristics of VR—visualization, traceability, immersion, and interactivity—map directly to the four-dimensional engagement framework: visualization supports cognitive engagement, traceability supports behavioral engagement, immersion enhances emotional engagement, and interactivity promotes active engagement. The Cognitive Affective Model of Immersive Learning (CAMIL) model further clarifies the interplay among cognitive, affective, and motivational effects in immersive learning [31].

For laboratory safety training, VR's advantages are particularly relevant. It enables learners to understand spatial layouts, identify hazards, rehearse emergency procedures, and practice high-risk tasks safely. Three-dimensional visualization reduces cognitive load, situational simulations enhance motivation, and interactive exploration supports repeated practice, addressing the deficiencies of traditional training. When combined with constructivist and experiential learning theories, VR can significantly enhance learning outcomes [32]. VR also allows remote, on-demand access, reducing teaching costs and improving educational equity in line with Open Educational Resources principles [33,34].

### *2.3. Application of Virtual Reality in Education*

The use of VR in education has expanded across disciplines, especially in areas requiring practical skills, spatial understanding, and risk management. In medical education, VR supports learning of organ structures, surgical procedures, and nursing skills, providing safe alternatives for high-risk operations and improving skill transfer [35,36]. In science courses, VR enables simulation of chemical reactions and physical phenomena, reducing material consumption, lowering costs, and supporting repeatable practice for better skill acquisition [37-39].

In architecture and construction, VR's immersive spatial representation helps learners understand complex layouts, support interior design, and analyze urban spaces. Integration with building information modeling further enables real-time design collaboration [40]. In engineering and construction education, VR facilitates equipment operation simulation, construction process training, and industrial safety exercises, providing repeated practice in a risk-free environment and enhancing learner engagement and safety awareness [41].

Overall, VR has demonstrated unique value in multiple educational fields, particularly in spatial visualization, scenario simulation, and learner engagement, providing both theoretical and technical foundations for its application in engineering education and laboratory safety training.

### *2.4. Point Cloud Optimization*

The large data volume of a point cloud often limits its broader applications [42]. In particular, under networked environments, strict bandwidth constraints may lead to data loss [43]. Therefore, point cloud optimization has become an important technical issue.

As one commonly employed optimization approach, point cloud simplification aims to reduce the number of points while preserving key information. Methods can generally be divided into two categories: global simplification and local simplification [44]. The former mainly relies on global point sampling strategies, such as grid-based and voxel-based subsampling, which offer high computational efficiency and relatively uniform sampling distributions. The latter places greater emphasis on the preferential preservation of local features, such as Gaussian curvature, in order to improve methodological flexibility.

By contrast, point cloud compression is another widely considered optimization approach, which focuses more on reducing data volume by exploiting the spatial correlation among points [42,44,45]. According to the standardization frameworks released by Motion Picture Experts Group (MPEG), point cloud compression can be divided into Video-based Point Cloud Compression (V-PCC) and Geometry-based Point Cloud Compression (G-PCC) [46]. The former maps three-dimensional point clouds into two-dimensional representations so that mature video coding frameworks can be used for compression [47]. The latter mainly relies on recursive three-dimensional spatial partitioning mechanisms such as octrees to perform geometry coding directly in 3D space [46].

Overall, point cloud simplification and point cloud compression share a common goal and can complement each other. For instance, through a combined optimization strategy of "simplification first, then compression", the resource demands of point clouds in storage, transmission, and processing can be further reduced [42]. For virtual reality applications, systems must balance scene realism with loading efficiency, making point cloud optimization particularly important. In recent years, machine learning and deep learning methods have also been gradually introduced into this field, providing new directions for improving the performance of related systems [48].

## 3. Study Site and Data

The development of a virtual engineering laboratory requires careful consideration of the physical environment to be digitized, the form of data used to represent it, and the procedures for acquiring and preparing that data. This section outlines the rationale for selecting the study site, explains the choice of point cloud data as the primary spatial representation, and details the workflow adopted for data acquisition and preprocessing.

### *3.1. Study Site Selection*

A large-scale engineering laboratory exceeding 4,000 m² was selected as the target environment for virtual reconstruction. This facility serves as the primary venue for core civil engineering teaching and experimental activities. As illustrated in Figure 1, the laboratory features a highly complex spatial configuration comprising seven distinct functional zones. Each zone accommodates specialized testing equipment, along with dedicated areas for material storage and specimen preparation. In addition, the laboratory is equipped with extensive safety infrastructure, including multiple emergency exits, fire-fighting systems, and first-aid facilities.

The combination of functional diversity, dense equipment arrangement, and comprehensive safety installations presents substantial challenges for spatial understanding and operational training. These characteristics make the laboratory an ideal candidate for investigating spatial cognition, safety training, and facility familiarization within a virtual environment.

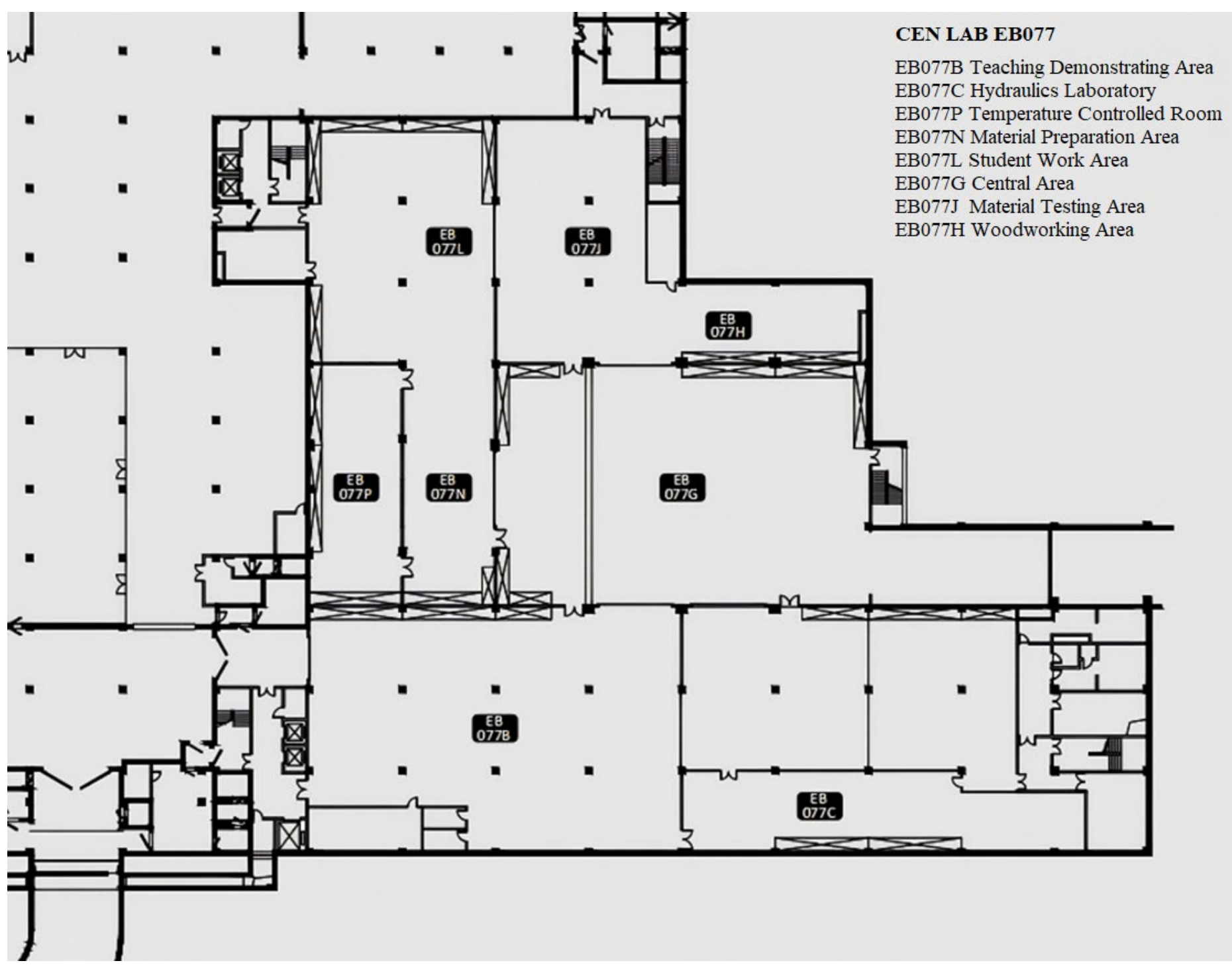

**Figure 1.** Overall layout of the considered civil engineering laboratory.

*3.2. Selection of Data Representation*

The choice of data representation plays a decisive role in determining the realism, accuracy, and pedagogical effectiveness of a virtual environment. Common approaches to three-dimensional scene reconstruction include manual 3D modeling, panoramic or image-based methods, and laser scanning. In this study, point cloud–based digital twin technology was selected to represent the laboratory environment for several reasons.

Compared with manual 3D modeling, laser scanning enables rapid and direct acquisition of high-resolution spatial data, accurately capturing complex geometries and fine details while avoiding the time-consuming modeling process and the potential loss of geometric fidelity [49]. This capability is particularly critical for engineering laboratories, where dense equipment layouts and intricate structures demand millimeter-level accuracy. Moreover, point cloud datasets allow localized rescanning and incremental updates, thereby supporting efficient long-term maintenance and adaptation to changes in laboratory layouts.

In contrast to image-based representations, point cloud data inherently preserve three-dimensional spatial information, enabling precise measurement and analysis of spatial relationships [50]. This supports real-scale visualization of distances, proportions, and orientations within the virtual environment, which is essential for reducing cognitive load during navigation, safety training, and operational tasks. In addition, point cloud can be colorized using aligned imagery data through data fusion. Collectively, these advantages motivated the adoption of point cloud data as the most suitable representation for the large-scale and spatially complex civil engineering laboratory considered in this study.

*3.3. Data Acquisition and Preprocessing*

Prior to data collection, an on-site survey was conducted to identify key areas and facilities that required emphasis in the virtual training environment. These included major experimental zones, clusters of critical equipment, and safety-related infrastructure such as fire extinguishers, first-aid kits, and emergency exits. Based on the laboratory's spatial layout and anticipated occlusion conditions, scanning stations were strategically planned to ensure complete spatial coverage, clear visibility of critical elements from individual stations, logic routes for space navigation, and sufficient overlap between adjacent scans to support robust point cloud registration [51,52]. These considerations led to a total of 22 pre-defined scan stations for subsequent laser scanning. These selected scan stations supported the subsequent establishment of waypoints in the developed virtual system.

**Table 1.** Summary of the Specifications of Leica RTC360 Scanner and Data Acquisition Parameters

| Category | Parameter | Value / Description |
|---|---|---|
| Scanner Specifications | Measuring rate | Up to 2,000,000 points per second |
| | Field of view | 360°(horizontal) × 300°(vertical) |
| | Selected resolution | 3 mm or 6 mm at 10 m |
| | Range | 0.5 – 130 m |
| | Accuracy | 1.0 mm + 10 ppm for range accuracy<br>18″ for angular accuracy |
| | Integrated cameras | 36 MP 3-camera system (0.4–0.76 μm in spectral range) |
| | Positioning system | Integrated Visual Inertial System (VIS) |
| Data Acquisition | Number of scan stations | 22 scan stations |
| | Type of collected data | Point cloud and RGB images (fused in Leica Cyclone) |
| Registration | Registration method | Preliminary real-time alignment between scans with the scanner's VIS; Cloud-to-cloud fine registration using Leica Cyclone |
| | Registration accuracy | Several millimeters in root mean squared error for most paired scans; slightly over 10 millimeters for a few paired scans |

Data acquisition was carried out outside peak laboratory usage hours to minimize noise and artifacts caused by human movement. Both point cloud data and imagery data were collected using a Leica RTC360 scanner. The scanner specifications are summarized in Table 1. During data acquisition, the scanner's integrated Visual Inertial System (VIS) was enabled to support real-time alignment between multiple scan setups. The acquired individual point clouds were first preliminarily registered using the scanner's VIS and subsequently refined through the cloud-to-cloud alignment in Leica Cyclone to integrate all point clouds into a unified coordinate system. Registration accuracy was evaluated using the root mean squared error, which was within 10 millimeters for most paired scans but slightly exceeded 10 millimeters for several paired scans, as reported in Leica Cyclone. In addition, point clouds were also fused with synchronously captured RGB image information in Leica Cyclone to generate a colorized point cloud.

The resulting raw dataset amounted to approximately 17.2 GB, exceeding the practical limits for real-time WebVR visualization. To address this constraint, point cloud simplification was conducted using a global distance-based downsampling strategy (a minimum point spacing of 5 mm). This process reduced the dataset from several billion points to approximately 300 million points (approximately 5.8 GB). This level of data simplification did not normally cause loss of essential geometric detail and visual fidelity [53,54]. Because laser scanning typically generates higher point densities near the scanner and overlapping observations of the same surfaces from multiple stations, the downsampling procedure also mitigated redundancy and produced a more uniform point distribution. This optimized dataset was therefore better suited for subsequent WebVR-based rendering and interactive applications.

*4. Virtual Laboratory Development*

To construct the virtual engineering laboratory, we developed a hybrid Unity–Potree WebGL approach. In this architecture, Potree handles efficient browser-based rendering of large-scale point cloud data, while Unity is responsible for interaction logic, user interface design, and educational functionalities. Both components are deployed using WebGL technologies, enabling seamless integration within a web environment.

As illustrated in Figure 2, the overall development workflow consists of four main stages: (1) determining an appropriate VR presentation format, (2) evaluating and choosing a suitable point cloud rendering platform, (3) designing and implementing interactive educational functions in Unity, and (4) integrating all system components. Through iterative prototyping and optimization, the system was progressively refined to ensure sufficient visual fidelity, interaction quality, and usability for laboratory training. The following sections describe the development process in detail and present the final virtual laboratory system.

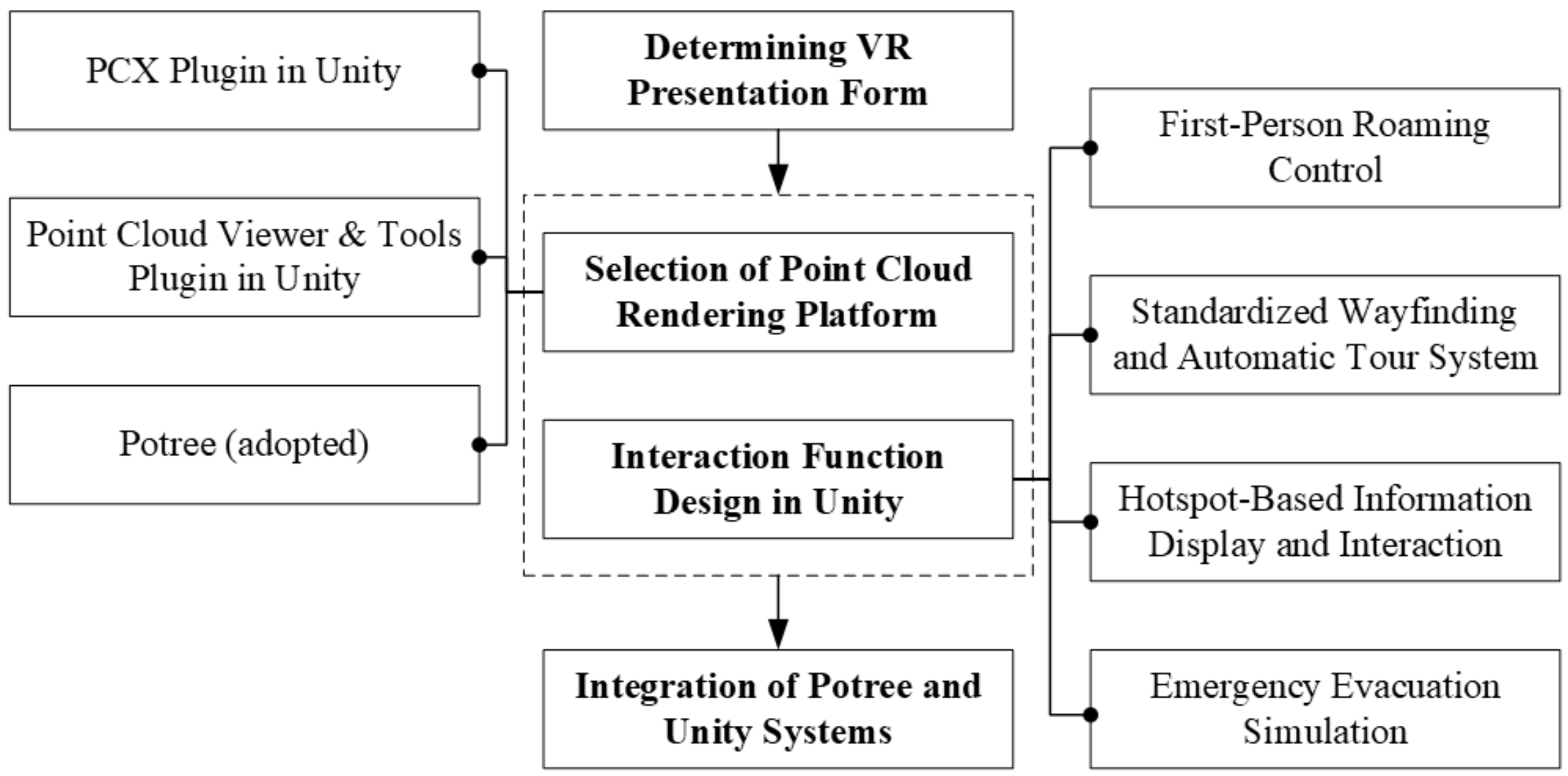

**Figure 2.** Overall methodology of developing the virtual laboratory system.

*4.1. Determining VR Presentation Form*

VR technologies can represent real-world environments at varying levels of immersion [55]. While fully immersive VR headsets can enhance sensory engagement, prior studies have shown that excessive immersion may hinder learning outcomes in domains such as spatial cognition and safety training due to cognitive and emotional overload [56]. In such cases, heightened sensory stimulation may distract learners from core instructional objectives, underscoring the importance of selecting an appropriate level of immersion.

In this study, WebVR was adopted as the presentation format. Built on open web technologies such as WebGL, WebVR offers strong cross-platform compatibility and can be accessed directly through standard web browsers. This allows the virtual laboratory to be deployed across classrooms, laboratories, and personal computers without specialized hardware such as VR headsets. In addition, WebVR supports simultaneous multi-user access, eliminates the need for dedicated physical spaces or complex installations, and lowers the operational threshold for both instructors and learners [57].

From an educational perspective, WebVR also reduces the likelihood of VR-induced discomfort, thereby supporting sustained learner attention. Its ability to be embedded within existing learning management systems or accessed via a Uniform Resource Locator (URL) further facilitates integration into established instructional workflows. Collectively, these advantages make WebVR well suited to the pedagogical objectives of engineering laboratory training.

*4.2. Selection of Point Cloud Rendering Platform*

A key technical challenge in developing the point-cloud-based virtual laboratory was the efficient loading and rendering of large-scale, high-density point cloud data. Although dense point clouds improve geometric realism, they place substantial demands on GPU, CPU, and memory resources, which can lead to reduced frame rates or system instability.

To address this challenge, a level-of-detail (LOD) strategy was employed. The point cloud was spatially partitioned into sub-blocks, and rendering resolution was dynamically adjusted according to the viewer's distance and field of view. Specifically, distant regions were rendered using lower-density representations, nearby regions with higher-density data, and non-visible regions were excluded from rendering. This approach effectively balances visual detail and computational performance [58].

Several rendering solutions were evaluated. Unity combined with the PCX plugin performed adequately for medium-scale datasets but exhibited limitations with larger point clouds, including long loading times, frame rate drops, high memory consumption, restricted point-size control, and WebGL memory constraints. Another Unity-based solution, Point Cloud Viewer & Tools, offered improved GPU acceleration but lacked WebGL support, necessitating substantial modifications to the development pipeline and significantly increasing implementation complexity and cost.

Based on a comparative evaluation of performance and deployability (Table 2), Potree was selected as the rendering platform. Potree is specifically designed for web-based visualization of large-scale point clouds and employs an octree-based, multi-level LOD mechanism [59]. This architecture enables stable browser-based rendering of the hundreds of millions of points required in this study.

To support Potree-based rendering, the original point cloud data underwent format conversion and hierarchical restructuring. First, CloudCompare was used to convert the E57 (*.e57) files into LAS/LAZ formats compatible with Potree. The converted data were then processed using PotreeConverter via the system terminal, which partitioned the dataset into a multi-level octree structure to enable dynamic LOD loading during runtime. This process involved iterative parameter tuning, such as adaptive point size and point budget, to balance visual fidelity and loading performance.

The final output comprised a self-contained web directory containing the HTML entry file (lab.html), associated JavaScript and CSS libraries, and core rendering data (e.g., metadata.json and octree.bin) stored in designated folders. This directory can be deployed directly on local or cloud servers without additional configuration, supporting flexible access and scalable deployment of the virtual laboratory.

**Table 2.** Comparison of Point Cloud Rendering Solutions.

| Options | Strength | Weakness |
|---|---|---|
| PCX Plugin in Unity | Simple to implement, easy to load, and supports WebGL. | Average visual effects, and insufficient capacity to handle large-scale data |
| Point Cloud Viewer & Tools Plugin in Unity | Enhanced memory management and LOD control | Deployment is complex and does not support WebGL. |
| Potree (adopted in this study) | High-performance point cloud rendering and LOD management, and compatibility with WebGL. | Limited native interaction functions, such as first-person navigation. |

### *4.3. Interactive Function Design in Unity*

While Potree excels at visualizing large-scale point clouds, its native interaction capabilities are limited. To enable spatial exploration and educational interaction, interactive modules were developed in Unity, which provides a mature interaction framework and a visual editor [60].

#### 4.3.1. First-Person Roaming Control

First-person spatial exploration constitutes the foundational interaction paradigm of the virtual laboratory. Prior studies indicate that the integration of visual perception and proprioceptive feedback facilitates accurate spatial encoding and enhances users' sense of presence [61]. To support this interaction mode, a First-Person Controller (FPC) plugin from the Unity Asset Store was adopted (Figure 3a).

The FPC represents the user through a gravity-enabled capsule collider coupled with a simplified ground collider to ensure stable and natural locomotion. A camera positioned at the top of the capsule provides an eye-level viewpoint consistent with human perception. Navigation is achieved via keyboard-based movement and mouse-driven view rotation. Key parameters, including eye height, translational speed, and rotational sensitivity, were systematically calibrated to approximate real-world laboratory navigation, thereby promoting spatial cognition, immersion, and active engagement.

#### 4.3.2. Standardized Wayfinding and Automatic Tour System

Given the spatial complexity of the laboratory environment, unrestricted free exploration alone may be insufficient for first-time users to develop accurate spatial orientation. To mitigate this limitation, a standardized waypoint-based wayfinding and automatic tour system was implemented.

As illustrated in Figure 3b, a set of waypoints was distributed throughout the virtual laboratory. These waypoints correspond to the scanner positions used during point cloud data acquisition, given that scanner positions were carefully determined to represent key laboratory locations, including major equipment areas, primary circulation paths, and designated safety exits, as presented in Section 3.3. Each waypoint encodes predefined spatial coordinates, camera orientation parameters, and sequencing information, providing a structured basis for navigation.

Two complementary navigation modes are supported. In the first mode, users may select a predefined waypoint from the navigation control menu, triggering instantaneous teleportation of the first-person controller to the corresponding location. This mode enables efficient and targeted exploration.

In the second mode, individual waypoints are connected into a continuous navigation path (Figure 3c), which is smoothed using curve interpolation. When activated, the camera automatically traverses this path at an optimized speed, providing a guided tour of the laboratory. This automatic tour mode is particularly effective during initial exposure, as it reduces cognitive load and allows users to acquire a holistic understanding of the laboratory layout without the demands of active navigation.

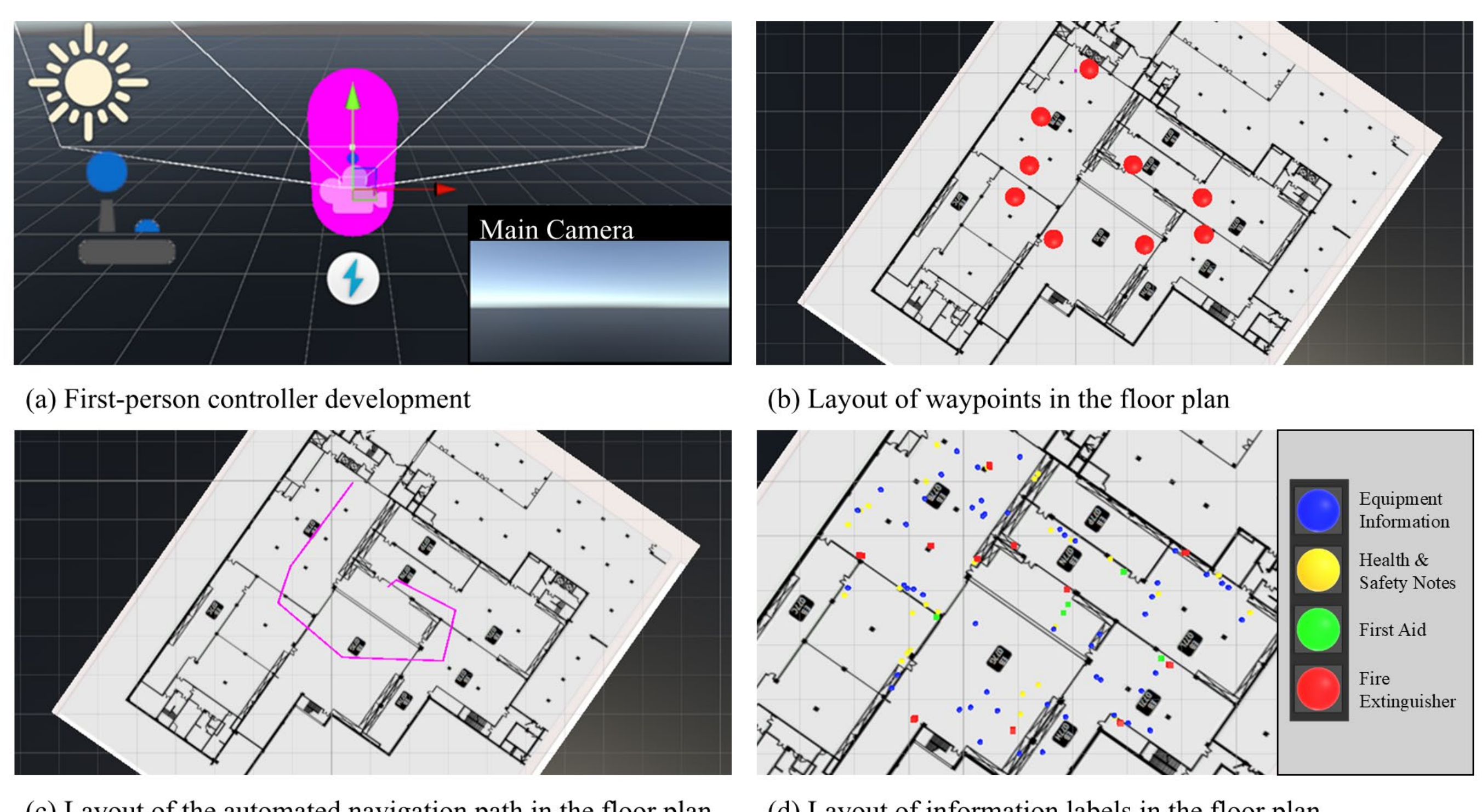

(a) First-person controller development

(b) Layout of waypoints in the floor plan

(c) Layout of the automated navigation path in the floor plan

(d) Layout of information labels in the floor plan

**Figure 3.** Interactive function developments.

4.3.3. Hotspot-Based Information Display and Interaction

Effective WebVR learning environments often rely on contextual annotations to support cognitive processing and long-term knowledge retention [21]. To facilitate efficient comprehension of laboratory equipment and safety infrastructure, a hotspot-based information and interaction system was developed using a lightweight, data-driven architecture built on Unity's ScriptableObject framework.

Hotspots were embedded at the locations of critical equipment and safety facilities, as shown in the plan view in Figure 3d. The operational logic of the hotspot system is presented in Figure 4. Each hotspot comprises a trigger collider, spatial metadata, and a unique identifier. Associated instructional content, including equipment descriptions, operating procedures, safety warnings, and illustrative images, is centrally managed through ScriptableObject assets. This approach supports efficient content reuse and batch updates.

When a user interacts with a hotspot, the system dynamically retrieves and displays the corresponding content in a pop-up interface, thereby supporting self-directed exploration and active knowledge construction within the virtual laboratory.

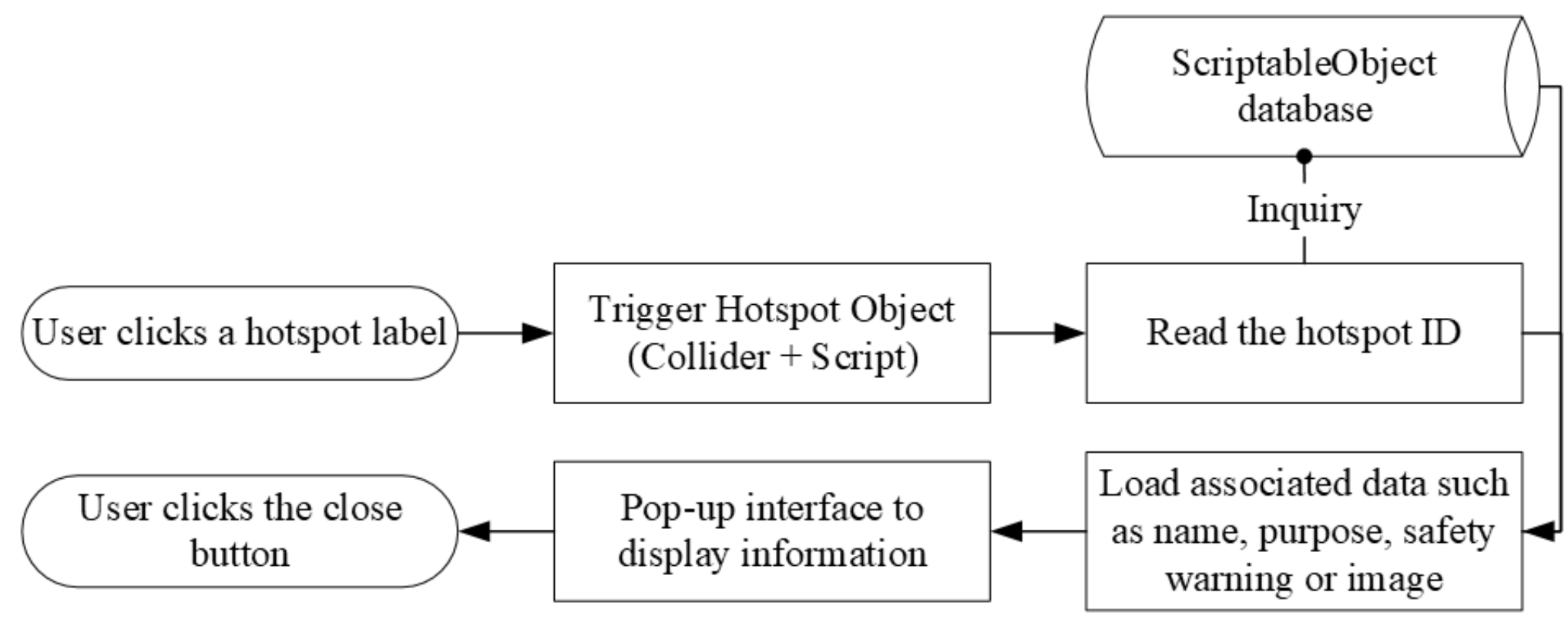

**Figure 4.** Operation logic of interactive hotspot labels.

4.3.4. Emergency Evacuation Simulation

To overcome the practical and logistical constraints associated with real-world safety drills, an emergency evacuation simulation module was integrated into the virtual laboratory. This module visually highlights safety exits and provides real-time, on-screen directional cues and distance indicators that guide users toward the nearest exit. The underlying mechanism of the evacuation simulation is illustrated in Figure 5.

Evacuation scenarios can be initiated from any location within the virtual environment, allowing users to navigate dynamically based on contextual guidance until a designated exit is reached. The risk-free nature of the virtual setting enables repeated practice without physical danger, thereby enhancing users' situational awareness, spatial decision-making, and emergency response capabilities under simulated conditions.

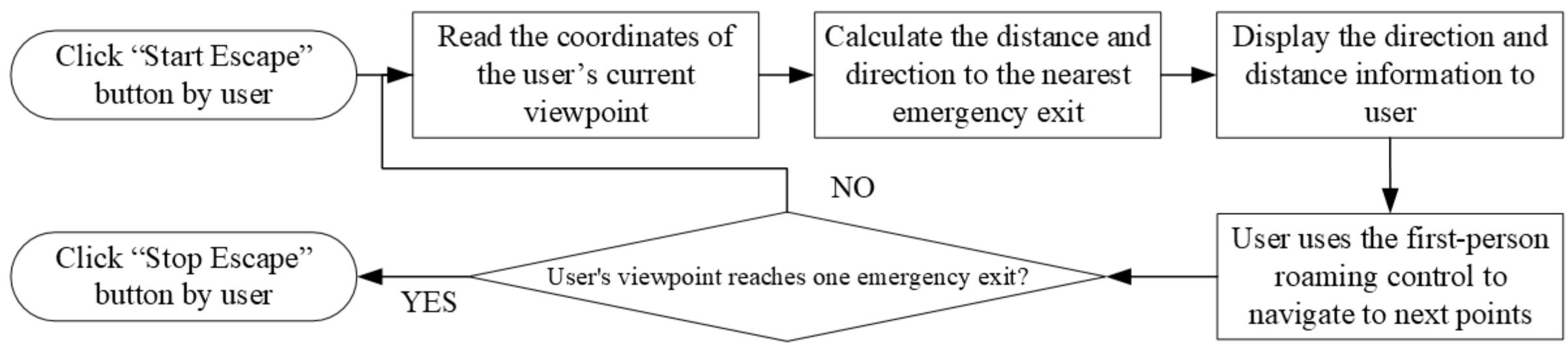

**Figure 5.** Operation logic of the emergency evacuation module.

*4.4. Integration of Potree and Unity Systems*

Although Unity and Potree belong to different technical ecosystems, both support WebGL, which enables their integration within a web-based environment. Each system is deployed as a web page and integrated using standard web technologies, including HTML, JavaScript, and CSS. In the composite interface, the Unity WebGL build functions as the background layer, while the Potree-rendered point cloud is embedded above it within an iframe. Visual stacking is controlled through HTML z-index settings, and synchronization between the two platforms is achieved via bidirectional communication between the iframe and JavaScript.

To ensure spatial alignment between Unity's interactive elements, such as hotspots and waypoints, and the Potree point cloud, a unified coordinate system is employed. Interactive elements are authored in Unity and exported as a JSON file containing unique identifiers and spatial coordinates. Potree loads this file to generate corresponding interactive hotspots within the point cloud scene. When a user selects a hotspot, Potree transmits the associated identifier to Unity through a JavaScript callback. Unity then queries its ScriptableObject repository and displays the relevant information panel (e.g., equipment descriptions or safety instructions), thereby establishing a closed-loop interaction mechanism between the two systems.

Input conflicts are avoided through a structured event-management strategy. The Unity Canvas is designated as the primary event receiver (pointer-events: auto), while the Potree viewport operates as a transparent overlay (pointer-events: none), with its native interaction features disabled. This configuration ensures that user navigation, first-person control, and interface interactions are consistently managed by Unity. Viewpoint control is likewise centralized in Unity, with its camera governing the overall perspective, thereby providing stable and coherent visual navigation, as Potree does not independently manage camera transformations.

Building on this integration approach, the virtual laboratory system architecture (Figure 6) is organized into three layers. The interface presentation layer comprises the Potree WebGL point cloud window and the Unity WebGL UI Canvas, together defining the final visual output. The interaction logic layer, driven by Unity, implements core functionalities such as user input processing, interaction triggering, navigation path computation, and automated tour playback. The data and content layer supplies shared resources for both systems, including point cloud data generated via PotreeConverter, hotspot definitions stored in JSON files and Unity ScriptableObjects, and navigation path configurations. Collectively, these layers constitute a hybrid virtual laboratory approach that provides a robust technical foundation for WebVR-based laboratory training.

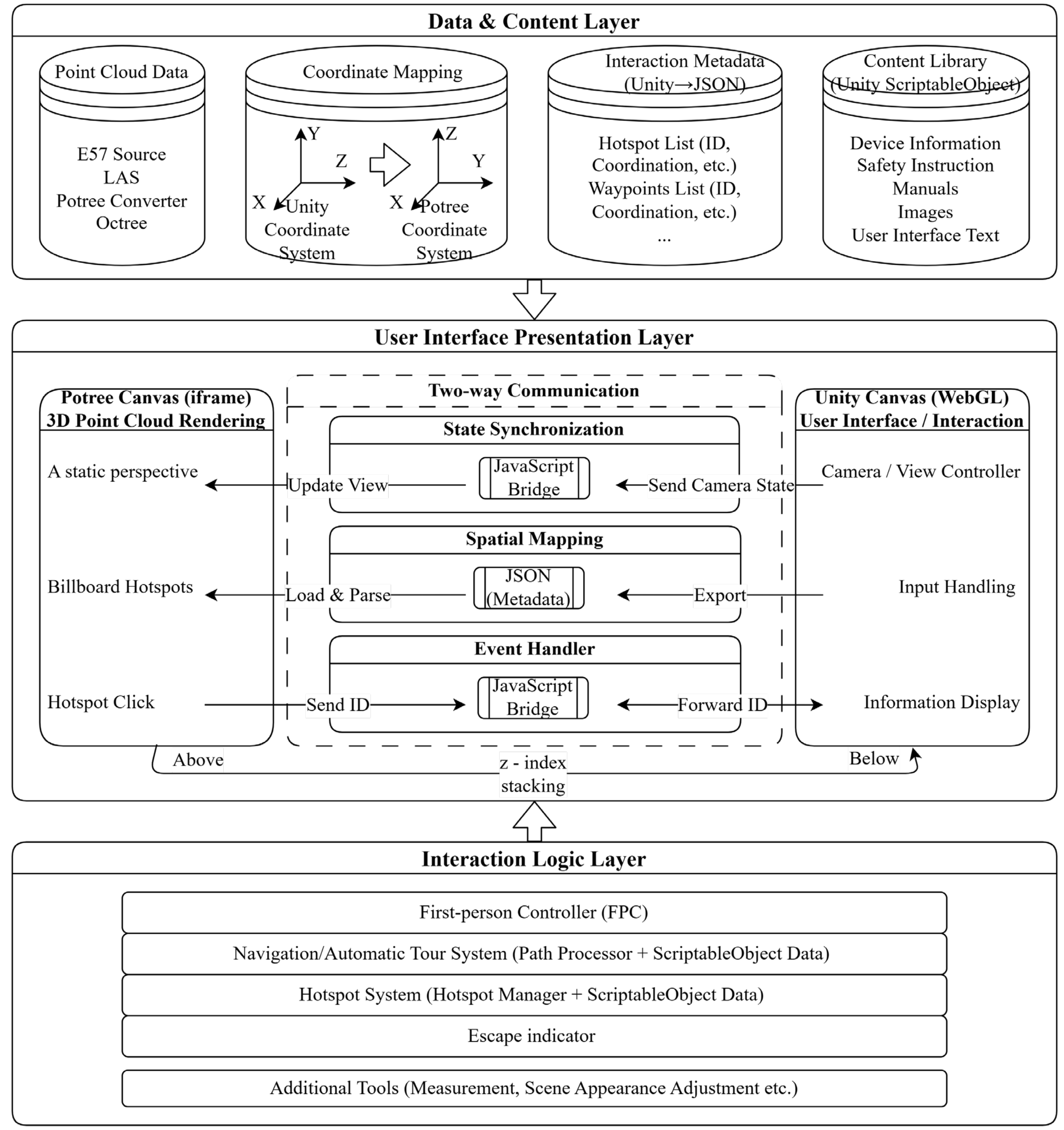

**Figure 6.** Hybrid Unity-Potree system confirmation.

### *4.5. Resulting Virtual Engineering Laboratory*

The completed virtual engineering laboratory is deployed as a web-based application and can be accessed directly through a standard web browser, without requiring specialized hardware or software installation. Upon entry, users are presented with the main interface (Figure 7), which provides essential laboratory information and navigation controls. The interface layout is designed to support efficient and intuitive interaction: the Potree toolbar is positioned on the left to provide rendering controls and measurement tools, the scene object list is located in the upper-right corner, and an emergency evacuation activation button is placed in the lower-right corner (Figure 7).

The system supports immersive first-person roaming, enabling users to freely explore the laboratory environment using standard keyboard and mouse inputs. Representative views captured from different viewpoints during first-person navigation are shown in Figure 8. For users who prefer a guided experience, an automatic tour mode can be activated via the "Auto-Tour" option (Figure 7). This mode moves the camera along a predefined route and can be paused or interrupted at any time. In addition, a set of predefined waypoints is available through the "Navigation" menu (Figure 7), allowing users to instantly travel to key locations and quickly gain an overview of the laboratory's spatial layout and organization.

Interactive hotspot functionality is implemented to deliver four categories of laboratory-related information: equipment descriptions (Info Points, Figure 7), fire extinguisher locations, first-aid facilities, and health and safety notices. These information layers can be selectively enabled through the "Scene Layers" panel, as described in Section 4.3.3. By interacting with icons embedded in the virtual environment, users can retrieve detailed content displayed in a pop-out information panel. An example of such content is shown in Figure 9. Viewed items are automatically logged and visually marked within the "Scene Layers" panel, enabling progress tracking within each category and supporting structured, goal-oriented exploration.

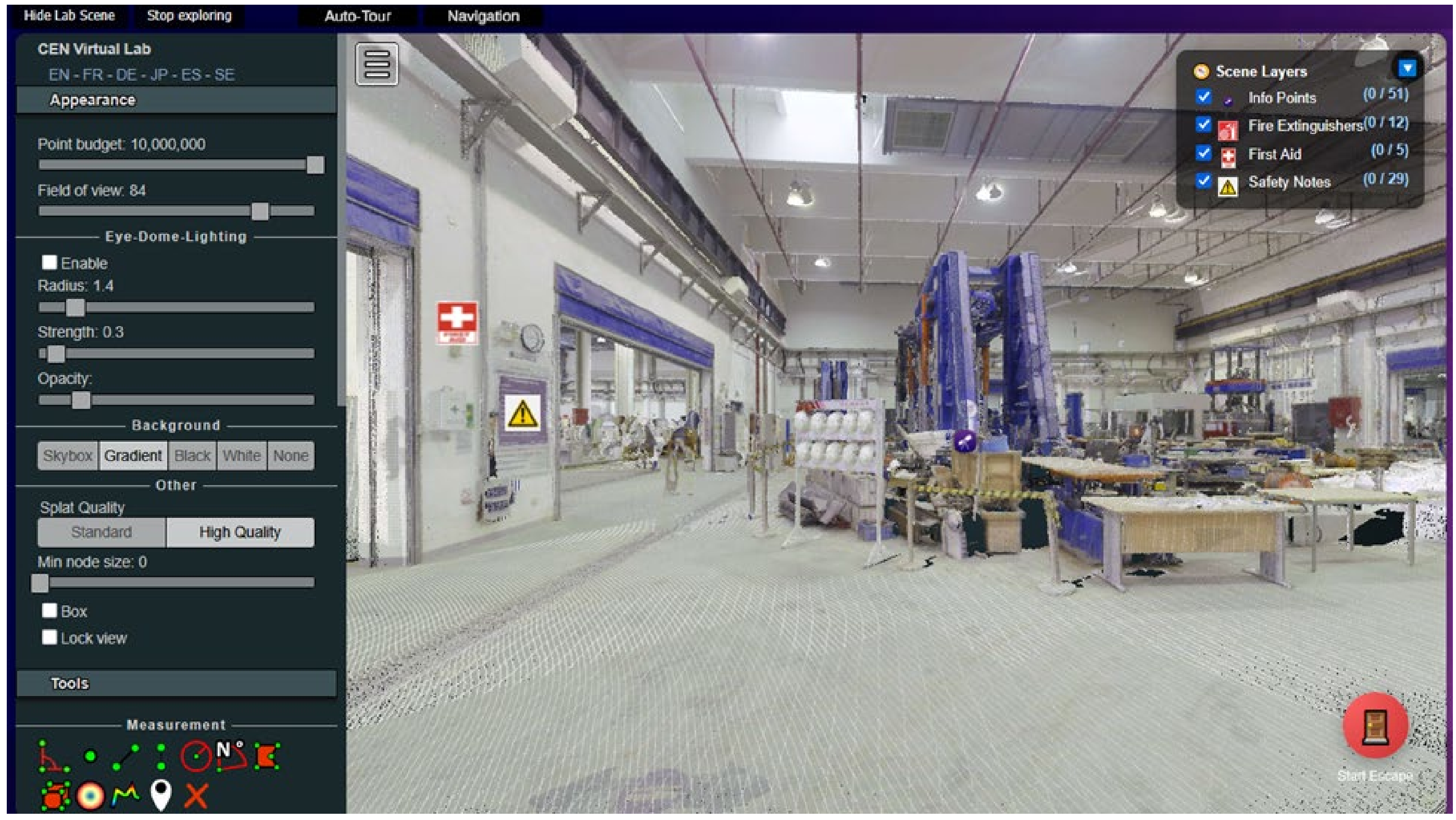

**Figure 7.** Virtual laboratory scene: "Scene Layers" shows all defined hotspots; "Auto Tour" enables automatic tour through a predefined route, "Navigation" switch between standardized viewpoints and first-person navigation, "Start Escape" enables emergency evacuation activity where users are guided to the nearest emergency exit.

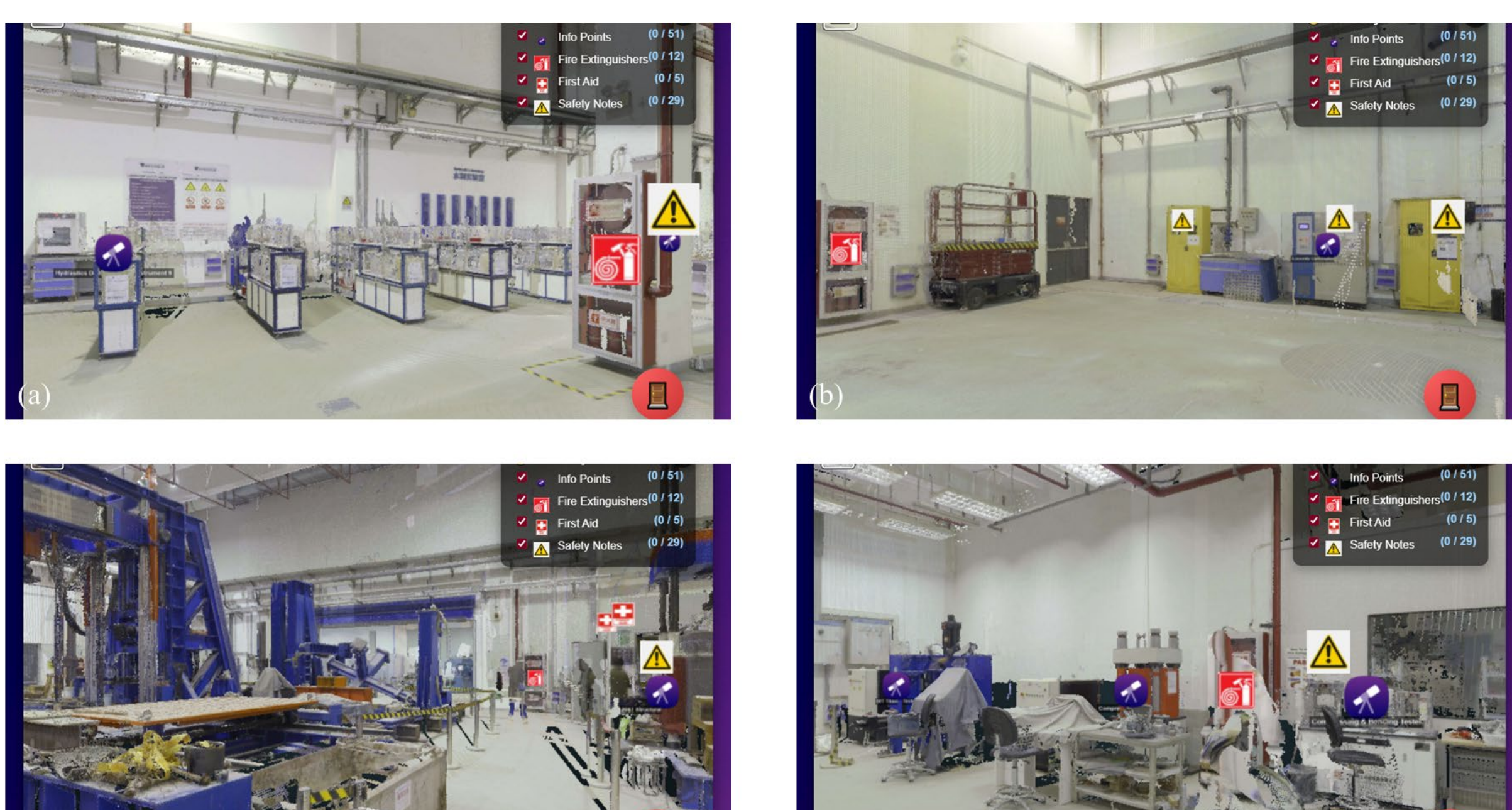

**Figure 8.** Examples of observed virtual scenes from various viewpoints using the first-person roaming: (a) hydraulic benches, (b) temperature control facilities, (c) 500T testing system, and (d) material testing area.

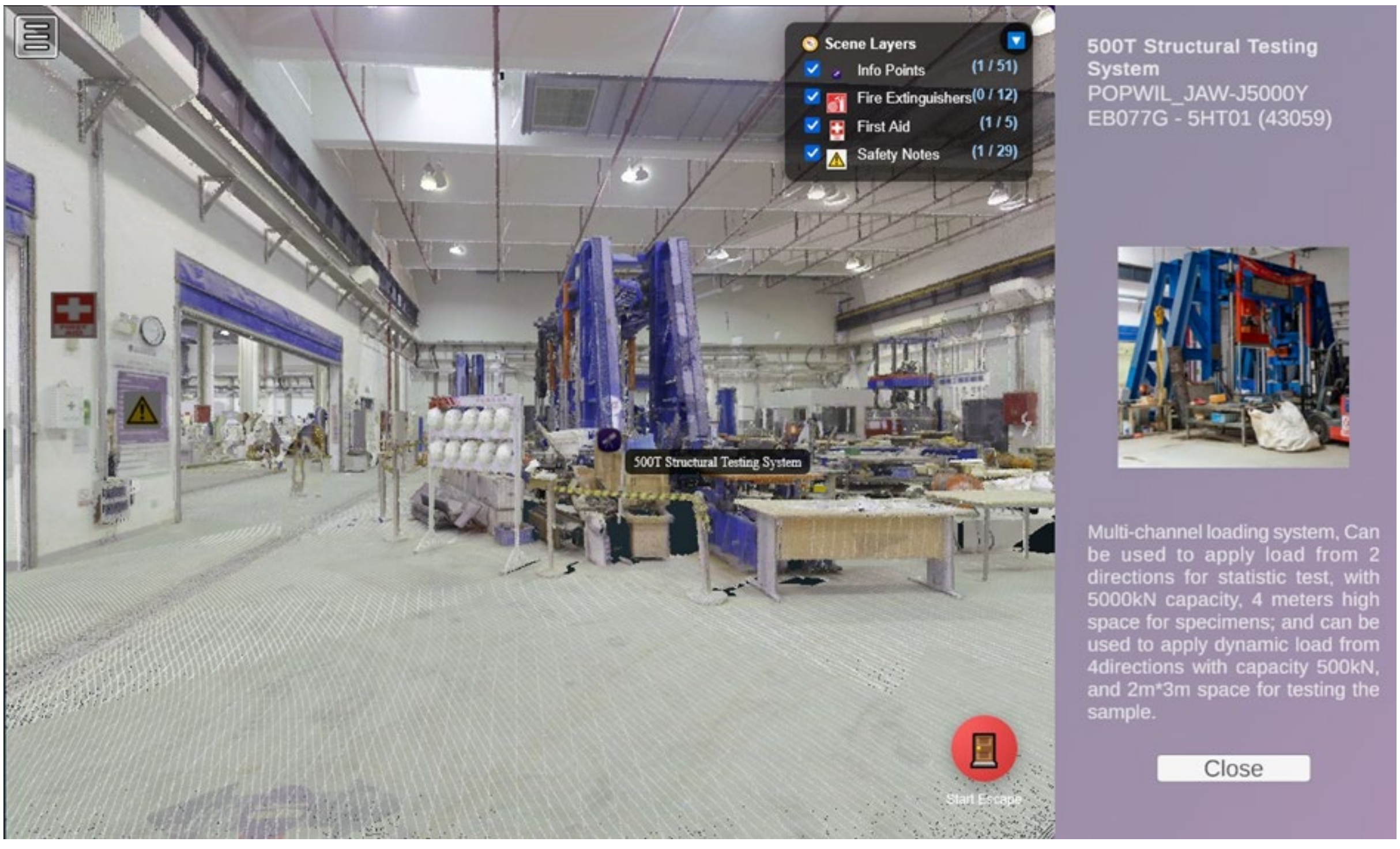

**Figure 9.** Example of retrieved equipment content at one equipment hotspot (i.e., Info Points), where 1/51 suggests one out of 51 equipment has been viewed.

To further enhance training effectiveness, the system includes an emergency evacuation simulation, activated via the "Start Escape" control. Once initiated, the system provides real-time, on-screen guidance that directs users toward the nearest safety exit. Successful completion of the evacuation procedure is confirmed by a visual indicator displayed at the exit location (Figure 10).

Beyond its instructional features, the virtual laboratory leverages Potree's built-in toolset, accessible from the left-hand toolbar. These tools allow users to adjust point cloud rendering parameters and perform spatial measurements, including coordinate querying, distance measurement, and area calculation. Such measurement capabilities represent a key advantage of point-cloud-based virtual environments over imagery-based WebVR systems and provide valuable geometric information for laboratory planning, spatial analysis, and pre-experimental preparation.

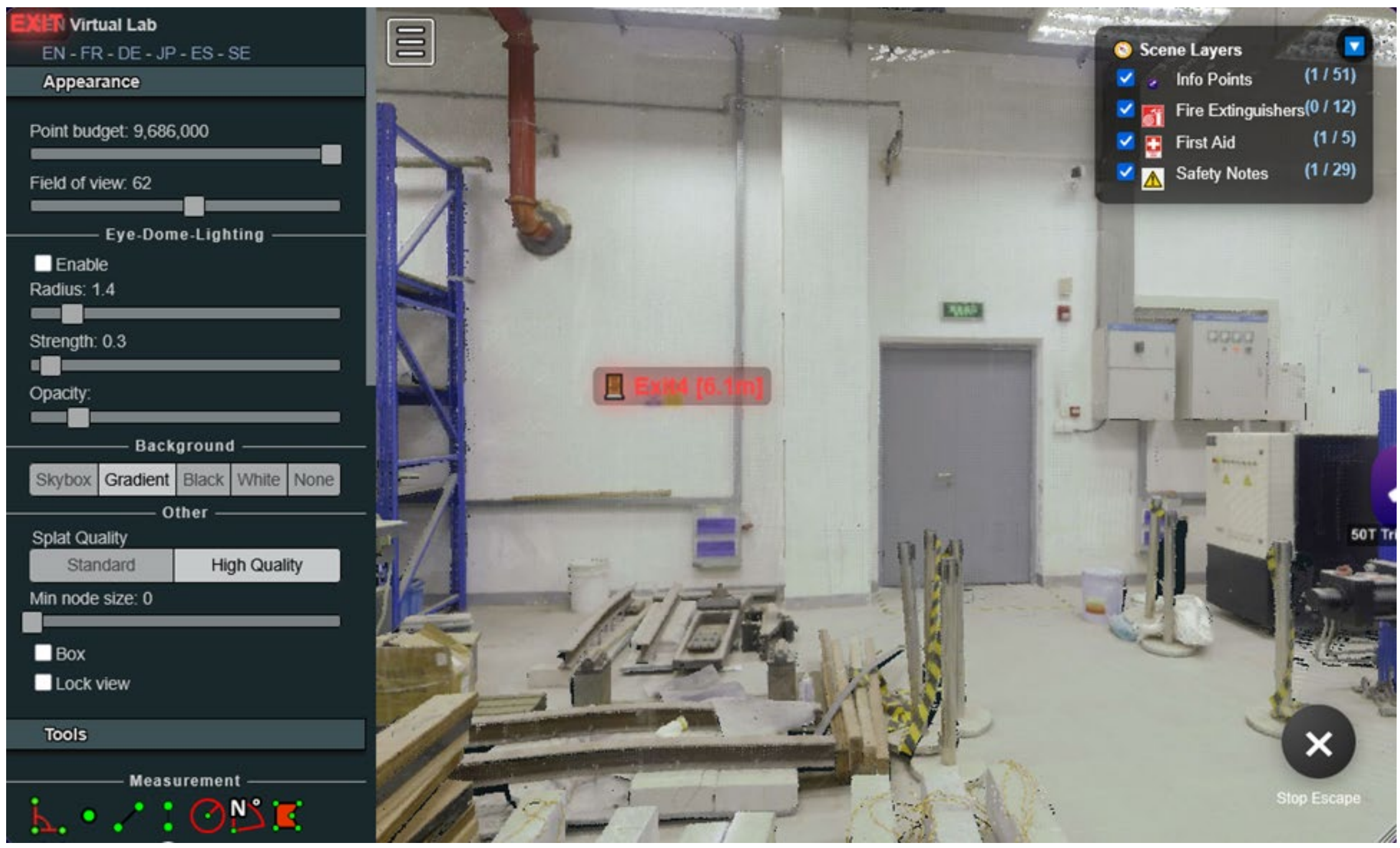

**Figure 10.** Example of the emergency evacuation simulation to one of the emergency exits, in which the distance and direction information is being constantly updated and displayed to users under the first-person roaming.

## 5. User Feedback Survey and Evaluation

A user survey was conducted to evaluate the virtual laboratory's usability, perceived educational effectiveness, and suitability for safety training and facility familiarization. This section presents the questionnaire design, survey administration procedures, and a comprehensive analysis of the survey results.

### *5.1. Questionnaire Design*

The questionnaire was developed to capture user perceptions in alignment with the instructional goals and functional characteristics of the virtual laboratory. It focused on three evaluation dimensions: system usability and interaction quality, including visual fidelity and navigation experience; system usefulness, with particular emphasis on spatial cognition, safety awareness, and training value; and overall user acceptance, including willingness to adopt the system for future learning or training activities. The questionnaire combined closed-ended and open-ended questions. Closed-ended items, presented in single-choice and multiple-choice formats, were used to collect demographic information and quantify user perceptions for subsequent statistical analysis. One open-ended question complemented these measures by allowing participants to elaborate on their experiences, identify issues, and propose suggestions for system improvement.

### *5.2. Participants and Survey Administration*

Survey participants comprised undergraduate students, postgraduate students, and laboratory staff members. Participation was entirely voluntary. The survey protocol was approved by the University Research Ethics Review Panel, and the survey was conducted in accordance with relevant guidelines and regulations. Informed consent was obtained from all participants.

Before entering the virtual environment, all participants received standardized preparatory training, which included written instructions as well as on-site demonstrations and users' preliminary trials of system operation. To ensure comprehensive system engagement, all participants were instructed to perform a structured system exploration comprising three sequential tasks: (1) Free exploration where participants freely navigated the virtual laboratory using first-person roaming, automatic tour, and navigation functions; (2) Interactive information browsing where participants explored hotspot-based interactive information by clicking on embedded tags associated with laboratory equipment and safety facilities; (3) Emergency evacuation simulation where participants activated the system's evacuation mode and completed an evacuation task by following on-screen directional and distance indicators from their current location to one safety exit.

The exploration phase lasted approximately 10 minutes on average (excluding the times taken for a preliminary exploration during the on-site preparatory training), which was sufficient for participants to develop an overall impression of the virtual system. Upon completion of the exploration, participants completed an anonymous online questionnaire administered through the Wenjuanxing platform.

### *5.3. Survey Results*

#### 5.3.1. Participant Profile

A total of 35 valid questionnaires were collected. All student participants had previously completed the in-house static laboratory training required for visiting the physical laboratory, which consisted of online written documentation covering operational procedures and emergency protocols. This prior experience enabled participants to make fair comparisons between the WebVR training and the documentation-based training. It also allowed them to meaningfully relate the virtual environment to real-world laboratory conditions, which is essential for experiential learning and reflective comparison as emphasized in TEL and experiential learning theory [18,23]. As shown in Figure 11a, postgraduate students constituted the majority of respondents, a group typically characterized by higher levels of self-regulated learning and cognitive engagement.

Participants' prior experience with VR or 3D simulation systems is summarized in Figure 11b. Approximately 20% of respondents reported no prior experience, while 74.3% had limited experience and only 5.7% extensive experience. Despite this uneven familiarity, subsequent usability ratings were uniformly positive, suggesting that the system successfully reduced barriers to entry. This aligns with TEL literature, which emphasizes that learning outcomes depend not on technology novelty but on how intuitively and accessibly it is implemented [14]. From an engagement perspective, the system appears to support behavioral engagement by enabling immediate interaction without extensive prior training [15-17].

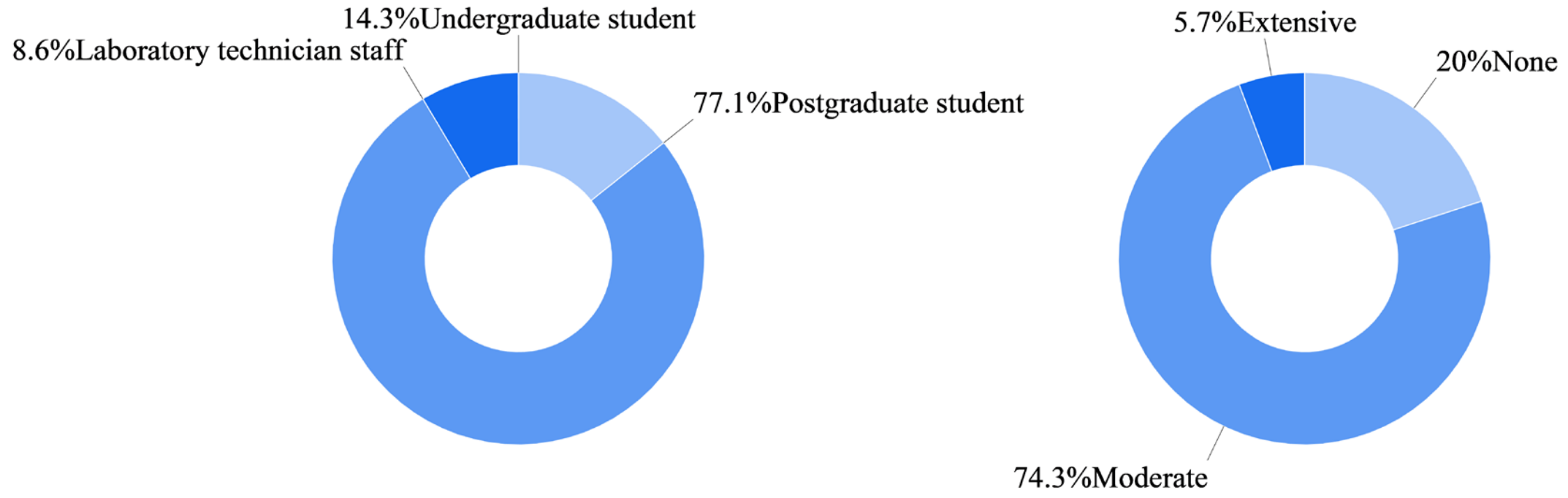

**Figure 11.** Participant demographic information: (a) Distribution of respondents' current roles, (b) Previous experience with VR or 3D simulation.

#### 5.3.2. Perceived System Usability and User Experience

Overall user experience with the system interface and navigation was consistently positive. As shown in Figure 12a, all participants rated system usability as either "excellent" (54.3%) or "good" (45.7%). This result is particularly significant given the participants' limited VR experience and supports the argument that well-designed immersive systems can lower extraneous cognitive load through intuitive interaction and clear information structuring [19-21].

Similarly, the system's visual quality received uniformly positive feedback (Figure 12a). High ratings for visual presentation directly relate to the role of visualization in supporting cognitive engagement. Dynamic, three-dimensional representations help learners integrate spatial, visual, and contextual information more effectively than static materials [21,22]. The positive evaluations therefore empirically support spatial learning theory, which posits that 3D environments enhance spatial perception and understanding in complex settings [29].

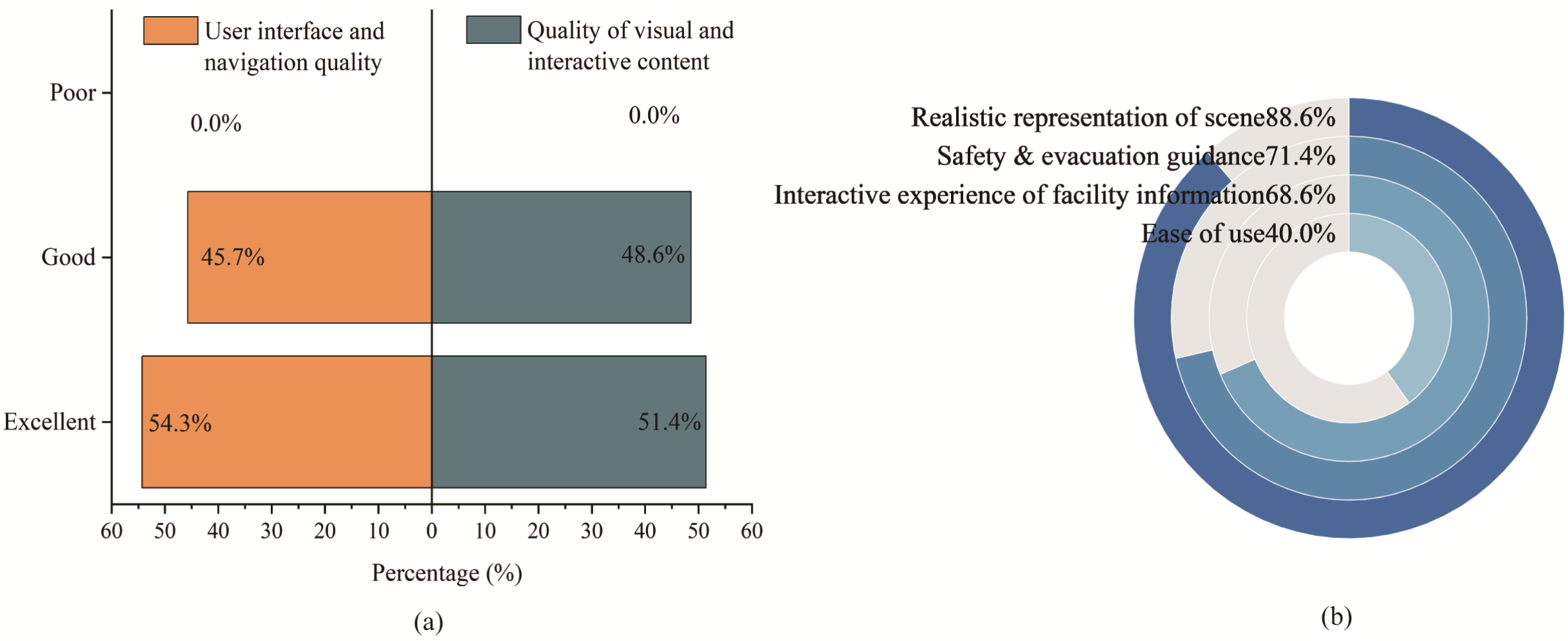

**Figure 12.** User evaluation of the virtual laboratory: (a) Overall ratings for quality of user interface and navigation, and quality of visual and interactive content, (b) User perceived strengths of the virtual laboratory.

### 5.3.3. Perceived System Strengths

When identifying perceived system strengths (Figure 12b), scene realism was selected most frequently (88.6%), followed by the practicality of safety and evacuation guidance (71.4%) and interactive information displays (68.6%). These findings strongly reflect the core affordances of VR, including immersion, visualization, and interactivity, which were identified in Section 2.2 as key drivers of learner engagement [28,30].

Scene realism contributes directly to emotional engagement by increasing presence and authenticity, consistent with CAMIL, which emphasizes the role of affect and motivation in immersive environments [29]. The high valuation of evacuation route visualization highlights the system's effectiveness in situated and experiential learning, enabling users to rehearse emergency scenarios safely, one of the principal advantages of VR for laboratory safety training [32,33].

In contrast, operational smoothness and ease of understanding were selected by only 40.0% of participants. This discrepancy suggests that while the system performs well in supporting cognitive and emotional engagement, aspects related to behavioral fluency (e.g., interaction smoothness) require improvement. This observation is consistent with TEL research cautioning that technological imperfections can hinder engagement if usability issues interrupt the learning flow [14].

### 5.3.4. Perceived System Usefulness in Supporting Learning

The virtual laboratory was widely perceived as engaging and motivating, supporting TEL arguments that learner engagement is central to educational effectiveness [13,14]. As shown in Figure 13a, all participants agreed that the system enhanced learning interest, with 57.1% rating it as "very attractive". This aligns with prior findings that immersive and interactive environments promote emotional engagement and intrinsic motivation by supporting autonomy and competence [15,16].

All respondents also agreed that the system improved their understanding of the laboratory's spatial layout (Figure 13b), with 74.3% rating navigation and spatial presentation as "extremely helpful". This result is consistent with spatial learning theory and VR research, which demonstrate that three-dimensional visualization enhances comprehension of complex spatial relationships while reducing extraneous cognitive load compared with static materials [19-22,29].

When compared with traditional online training methods (Figure 13c), 74.3% of participants reported a significant improvement in learning engagement. This finding reflects limitations of conventional online instruction in supporting behavioral and active engagement [18] and supports VR literature emphasizing "learning by doing" through immersive, exploratory interaction [30,32]. The virtual laboratory's immersive and interactive characteristics therefore effectively support multiple engagement dimensions defined by Reeve et al. [15,16].

Participants unanimously agreed that the virtual laboratory could effectively supplement traditional on-site training (Figure 13d), particularly for pre-visit familiarization and safety training. This perception aligns with prior studies identifying VR as a complementary tool for laboratory and engineering education, enabling repeated, risk-free practice and flexible access without replacing physical training [39,41].

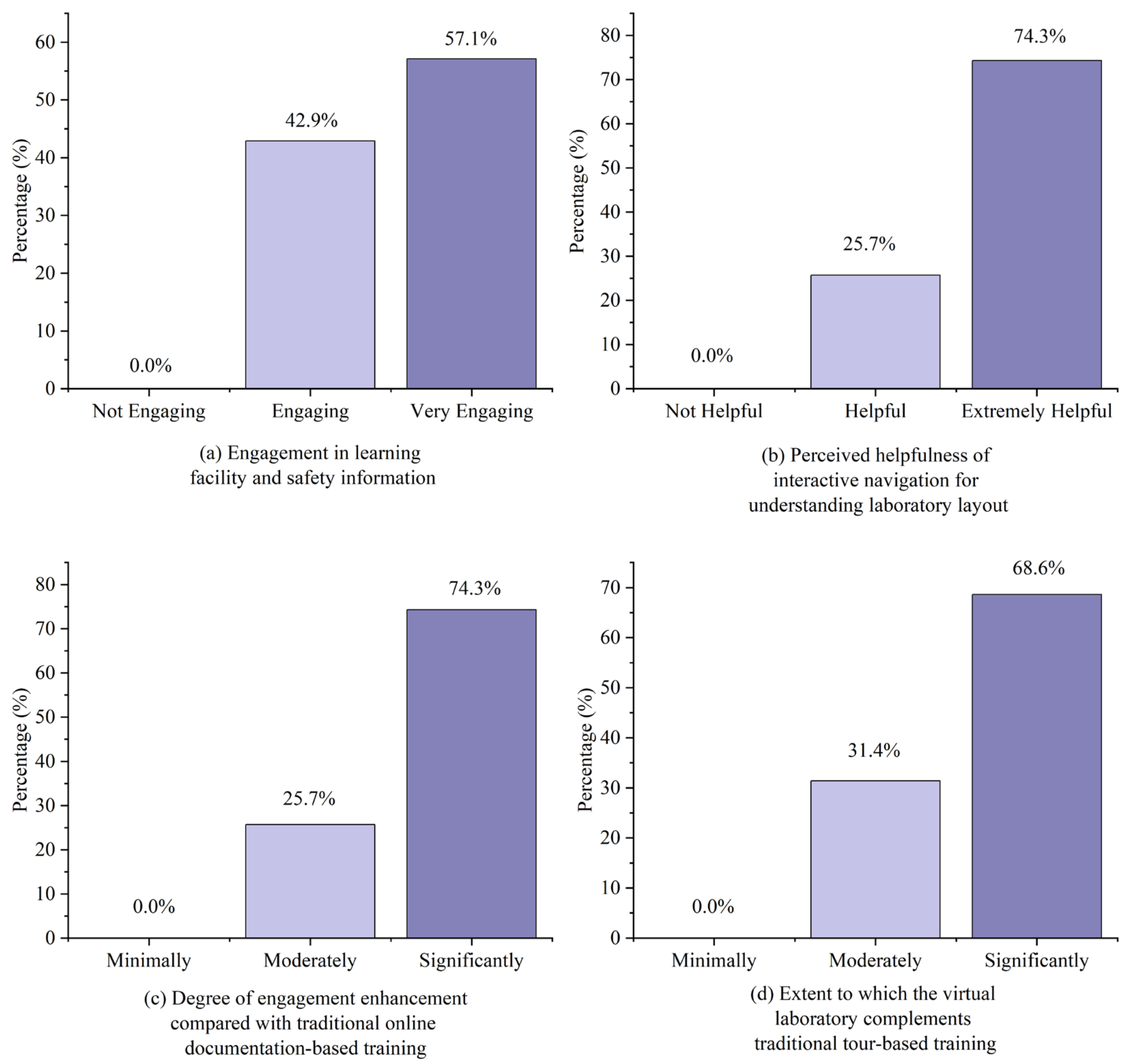

**Figure 13.** User perceptions of system usefulness and engagement within the virtual laboratory: (a) Engagement in learning facility and safety information, (b) Perceived helpfulness of interactive navigation for understanding laboratory layout, (c) Degree of engagement enhancement compared with traditional online training, (d) Extent to which the virtual laboratory complements traditional tour-based training.

5.3.5. Overall Acceptance and Recommendation

Overall acceptance of the system was high. As shown in Figure 14a, all participants expressed willingness to use the system in the future, with 68.6% indicating strong intention. Similarly, 82.9% of participants reported being "very willing" to recommend the system to others (Figure 14b). Such strong acceptance suggests that the system successfully supports active engagement, where learners perceive value in repeated use and peer recommendation.

These findings align with constructivist learning theory, which emphasizes that learners are more likely to internalize knowledge when they actively explore, manipulate, and reflect within meaningful contexts [23,25]. Moreover, high recommendation willingness echoes findings from prior VR education studies, which show that immersive, interactive systems enhance motivation and perceived learning effectiveness across disciplines [35,41].

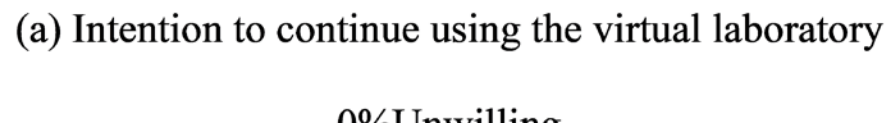

(b) Willingness to recommend the virtual laboratory to others

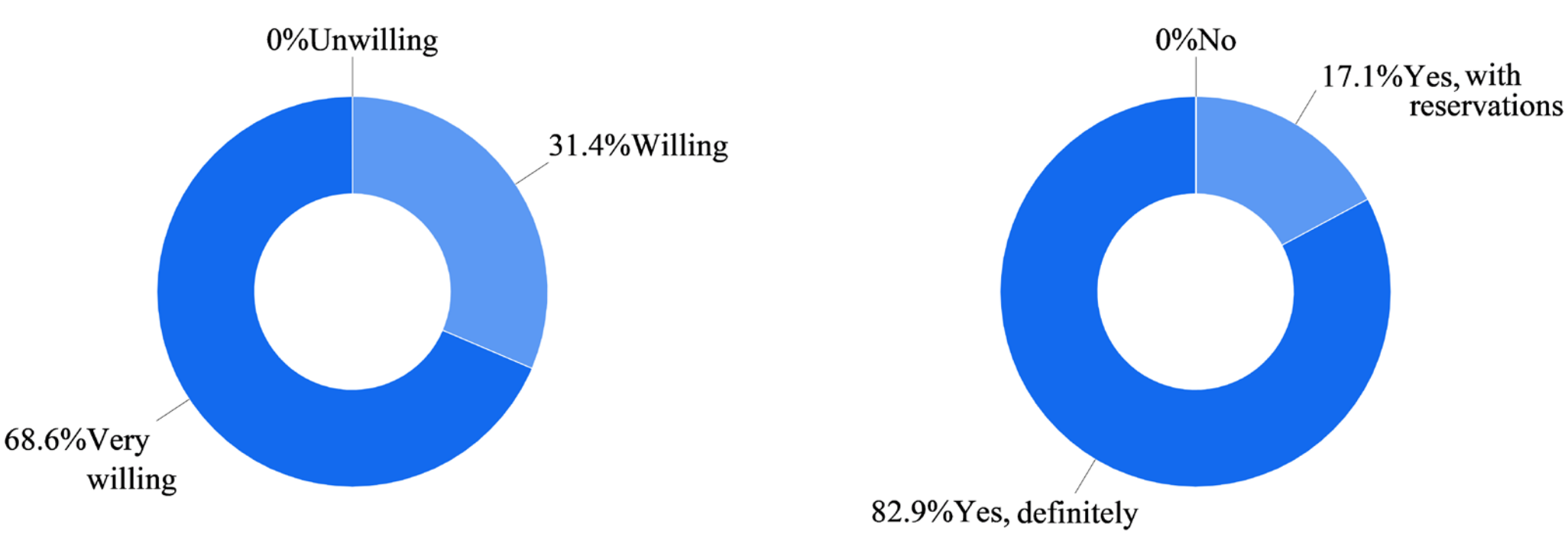

**Figure 14.** User acceptance and adoption intentions regarding the virtual laboratory: (a) Participants' intention to continue using the system for future study or work, (b) Participants' willingness to recommend the virtual laboratory to other students as a learning resource.

5.3.6. Thematic Insights from Open-Ended Feedback

Nineteen participants provided open-ended feedback in the survey. Thematic analysis of these responses reveals several recurring strengths and limitations that largely align with the quantitative findings and help clarify priorities for future improvement.

The most frequently mentioned concerns related to operational convenience and usability. Participants noted the lack of explicit on-screen guidance or instructional prompts, which increased the learning burden for first-time users. Several respondents also reported insufficient smoothness or responsiveness in certain interactions, including occasional lag and reduced sensitivity. Suggested improvements included expanding the camera rotation range to enable full 360° viewing and enriching information tags with deeper interactive content (such as videos or 3D models).

Feedback on visual quality focused on point-cloud artifacts and rendering performance. Participants reported issues such as gaps between points, occasionally slow rendering speeds, and limited sharpness. These observations indicate a need for further optimization of point-cloud preprocessing, level-of-detail management, and WebGL-based rendering performance.

Several respondents emphasized the importance of the emergency evacuation simulation as a core feature. Suggestions included using more prominent colors, clearer directional arrows, and stronger visual cues for escape guidance. Some participants noted that the current escape mode provided only direction and distance rather than explicitly shown evacuation routes, recommending clearer path visualization and obstacle-aware guidance.

Participants reported that the lack of collision detection and penetrable structural elements reduced realism. These responses suggest that users expect virtual laboratories to respect basic physical laws, indicating that collision components and boundary constraints should be incorporated in future iterations.

Despite these critiques, participants offered strong positive evaluations, praising the system's overall performance, educational value, clarity of equipment visualization, usefulness of the evacuation function, and creativity as a teaching and learning tool. These remarks reinforce the quantitative results and confirm the system's pedagogical relevance.

## 6. Discussion

### *6.1. Virtual System's Technical Strength and Limitation*

From a technical standpoint, the proposed WebVR system constitutes a substantive advance in reducing the barriers to deploying virtual laboratories. TEL research emphasizes that educational technologies must improve learning conditions while remaining feasible within institutional constraints [13,14]. By operating entirely within a standard web browser, the system minimizes hardware requirements and alleviates pressure on institutional teaching infrastructure, aligning with the broader educational informatization trend that prioritizes flexible, maintainable, and scalable digital resources [26,27]. Supportive feedback from technician staff highlights the practical viability of this approach for sustained instructional use, particularly in contexts where physical access to laboratories is limited.

The adaptation of colorized point cloud data enables the creation of a high-fidelity digital twin that faithfully captures the spatial complexity and context characteristics of a real engineering laboratory. Compared with manually modeled or image-based virtual environments, colored point clouds offer clear advantages, including rapid scene generation, efficient localized updates, and close alignment with real-world layouts. These characteristics are especially critical for applications in safety training and facility management.

At the same time, survey findings reveal inherent technical trade-offs associated with browser-based point cloud visualization. Although participants rated overall usability and visual realism positively, both quantitative scores and qualitative comments pointed to limitations in interaction fluidity, loading performance, and visual clarity. These issues stem primarily from the high density of point cloud data, the constraints of web-based rendering, and the lack of advanced physical interaction mechanisms. Taken together, the results suggest that while the system demonstrates functional and educational usability, further technical optimization is required to meet the higher expectations of experienced users and to support more demanding professional applications.

### *6.2. Educational Value and Learning Mechanisms*

The observed educational benefits of the virtual laboratory align closely with the TEL principles [13], which emphasize the improvement of learning conditions over technological novelty. By reconstructing the laboratory's spatial layout and equipment information using three-dimensional point clouds, the virtual system enables learners to perceive facilities and spatial relationships immersively, offering a level of contextual understanding that exceeds that of conventional instructional materials. This aligns with spatial learning theory, which argues that three-dimensional representations enhance learners' understanding of complex spatial relationships and improve retention [21,29].

A range of interactive features, such as first-person navigation, guided roaming, automated tours, and evacuation simulations, transform otherwise static laboratory content into an exploratory learning environment. This closely aligns with the CAMIL [31], which emphasizes the interplay between immersion, affective responses, and cognitive processing. Survey results indicate enhanced learner engagement across multiple dimensions, aligning well with Reeve et al.'s four-dimensional engagement framework [15]. Emotional engagement is supported by immersive first-person perspectives and high-fidelity visual scenes. Cognitive engagement is reinforced through accurate spatial representations and navigational aids, which facilitate scene comprehension while reducing cognitive load. In addition, interactive hotspots that provide equipment-specific explanations and safety guidance enable self-paced, repeatable learning, effectively addressing the limitations of time-constrained, instructor-led on-site instruction. Behavioral engagement is further promoted through hands-on exploration and task-oriented evacuation simulations. Notably, these benefits are consistently reported across different user groups, indicating that the system's educational value is not attributable to novelty effects but rather to its substantive learning affordances.

### *6.3. Limitations and Directions for Future Development*

Several limitations of the present study warrant consideration. From a technical perspective, the quality of point cloud data was constrained by acquisition strategies and hardware capabilities, leading to noise, artifacts, and occasional visual discontinuities. In addition, browser-based rendering also imposed constraints on loading stability and real-time performance. Future technical improvements should not only focus on the quality of the point cloud itself, but also address transmission efficiency in web-based deployment environments, for example by exploring more efficient point cloud compression strategies, on-demand streaming mechanisms, and Content Delivery Networks (CDNs) [43,46,47,62]. These approaches have the potential to reduce initial loading time and improve the scalability of the system under more challenging deployment scenarios.

Functionally, while navigation and evacuation features were positively received, users expressed expectations for more advanced capabilities, including obstacle-aware navigation, clearer evacuation paths, and more salient visual cues. Therefore, potential improvements include completion of occluded regions, more efficient data rendering strategies, the incorporation of collision effects to enhance physical realism, and more intelligent pathfinding algorithms capable of calculating optimal escape routes based on environmental obstacles. Meeting these expectations will require deeper integration of spatial data, rendering logic and scene semantics, an objective that remains challenging with currently available software tools WebVR and demands future development.

Our current virtual system visualizes the entire point cloud as a single integrated object and does not automatically distinguish individual facilities. Future research could incorporate semantic segmentation techniques to classify point cloud data corresponding to specific instruments or facilities. Such techniques have been widely applied to both point cloud data [63] and imagery data [64]. Integrating these approaches would enable users to more easily locate specific instruments during training and facilitate more intelligent facility management.

Although our virtual system is primarily designed for laboratory orientation, facility familiarization, and safety briefings, future work could expand interactive content to support task-based learning, such as equipment operation procedures, safety workflows, and multi-scenario simulations. More advanced extensions may involve AI-assisted virtual tutoring and automated safety hazard identification through the combined analysis of point cloud data and artificial intelligence techniques. In addition, deployment across distinct platforms, including immersive VR headsets, should be explored to balance accessibility with immersion.

Given that the primary objective of this study was to develop the technical approach for a hybrid WebVR system, the evaluation of system effectiveness was limited to overall measures of users' perceived usability, usefulness, and acceptance, rather than employing more comprehensive evaluation measures that are commonly used in system-level assessments. Nevertheless, a wider range of standardized evaluation instruments, such as the System Usability Scale (SUS), the Immersive Tendencies Questionnaire (ITQ), and the Simulator Sickness Questionnaire (SSQ), could be incorporated in future studies to enable more systematic assessment of usability, immersive experience, and potential discomfort. The inclusion of these additional metrics would also enhance the rigor and comparability of the research findings [65-67].

## 7. Conclusions

This study designed, implemented, and evaluated an immersive virtual engineering laboratory grounded in point-cloud–based digital twin technology and a Unity–Potree hybrid WebVR approach. A high-fidelity digital representation of a large, real-world engineering laboratory was constructed and made accessible entirely through a standard web browser. The virtual system integrates immersive navigation, interactive information hotspots for laboratory facilities and health and safety information, and emergency evacuation simulations. These features enhance learners' spatial understanding and safety awareness, while interactive information elements support knowledge acquisition, learning motivation, and sustained engagement. Methodologically, the study also demonstrates a practical and extensible workflow for integrating Unity and Potree, enabling efficient development of versatile, browser-based virtual scenes based on massive point cloud data.

Survey-based evaluation using an anonymous online questionnaire indicates that the proposed virtual system provided clear benefits in supporting learning. Invited users consistently affirmed the system's overall usability and usefulness, widely perceiving the virtual laboratory as a meaningful complement to traditional on-site instruction, particularly for pre-visit familiarization and laboratory safety training. At the same time, participant feedback identified concrete areas for improvement, including the treatment of occluded regions, level-of-detail optimization, rendering performance, and collision handling, thereby offering clear directions for future technical refinement.

Overall, these findings demonstrate that WebVR-based virtual laboratories have significant potential to enhance spatial understanding, equipment familiarity, and safety awareness. The proposed approach supports the broader adoption of technology-enhanced learning and offers a scalable, transferable model for virtual laboratory development in higher education.

**Acknowledgments:** The authors are grateful to all participants for completing the questionnaire sincerely.

**Data Availability Statement:** The data generated in this study is available from the corresponding author upon reasonable request.

**Institutional Review Board Statement:** The study protocol for the user survey was reviewed and approved by the University Research Ethics Review Panel. All procedures were conducted in accordance with applicable ethical guidelines and regulations.

**Informed Consent Statement:** Informed consent was obtained from all individual participants for participation in the survey. Anonymity of their identification and voluntary participation were assured. No identifiable data of participates were present within the manuscript.

**Conflicts of Interest:** The authors declare no competing interests.